%% file: ISSM_ariv.tex
\documentclass[%
 reprint,
superscriptaddress,
 amsmath,amssymb,
 aps,
prl,
twocolumn
]{revtex4-2}

\usepackage{graphicx,bm,braket,color,hyperref,comment}
\usepackage{fancybox}
\usepackage{ulem}
\hypersetup{colorlinks=true,citecolor=blue,linkcolor=blue,urlcolor=blue}

\newcommand{\bM}{ {\bm{M}} }
\newcommand{\bH}{ {\bm{H}} }

\begin{document}


\title{Inverse Single-sided Magnet
}

\author{Takayuki Ishitobi}
\affiliation{Advanced Science Research Center, Japan Atomic Energy Agency, Tokai, Ibaraki 319-1195, Japan}

\date{\today}

\begin{abstract}
A single-sided magnet generates a magnetic field on only one side while canceling it on the opposite side, a feature that has enabled diverse applications in both fundamental science and engineering. Here, we propose the {\it inverse} single-sided magnet: a non-ferromagnetic system that selectively attracts either the north or south pole of a ferromagnet while remaining unresponsive to the opposite pole. 
We demonstrate that such behavior can arise in microscopic octupolar magnets. To illustrate this, we analyze two minimal models: a coplanar magnetic structure with 120-degree ordering and a collinear magnetic structure with site-dependent anisotropy. In both cases, we find that the magnetization response is nonreciprocal with respect to the sign of the applied magnetic field. Notably, in the latter model, the system exhibits strong magnetization in one field direction and negligible response in the opposite direction. This diode-like behavior for magnetic fields suggests that inverse single-sided magnets could play a pivotal role in controlling magnetic interference, with potential impact comparable to conventional single-sided magnets.
\end{abstract}

\pacs{Valid PACS appear here}
\maketitle


{\it Introduction.---}A single-sided magnet exhibits the unique property of producing a magnetic field on one face while canceling it on the opposite side \cite{Mallinson1973-xc, Halbach1979-fb, Halbach1980-gi}. This asymmetric field configuration has enabled a wide range of applications, spanning fundamental research, industrial technologies, and everyday devices \cite{Halbach1985-sd, Zhu2001-ju, Coey2002-ht, Gieras2008-zh, Gieras2009-zb, Spaldin2010-qg}. In high-energy physics, single-sided magnets are essential for beam steering in particle accelerators; in transportation, they contribute to magnetic levitation systems. They are also widely employed in magnetic shielding, precision sensors, and everyday items like refrigerator magnets. The ability to spatially control magnetism enhances the efficiency of existing magnetic devices and opens avenues for nonreciprocal magnetic functionalities. 

This observation naturally raises the question: does the {\it inverse} of a single-sided magnet exist? While a conventional single-sided magnet attracts magnetic materials on only one side, we propose the {\it inverse single-sided magnet}--a non-ferromagnetic system that selectively attracts only one magnetic pole (either north or south) while remaining unresponsive to the opposite pole. Such a system would serve as a magnetic analog of a diode, potentially suppressing magnetic interference and enabling directional control in microscopic magnetic devices. In this study, we explore how this concept can be realized using simple theoretical models, aiming to uncover both underlying physical mechanisms and prospective applications. 

{\it Symmetry.---}To establish a symmetry-based framework, we revisit the case of conventional single-sided magnets. These are typically realized using a specific arrangement of permanent magnets known as the Halbach array \cite{Halbach1980-gi}. In such configurations, the balance between magnetic dipole and quadrupole moments plays a key role in generating a unidirectional magnetic field \cite{Mallinson1973-xc, Halbach1979-fb}. In contrast, we show that inverse single-sided magnets can arise from microscopic magnetic octupoles. This can be understood by expanding the magnetization response $\bM$ to an external magnetic field $\bH$ up to the second order: 
\begin{align}
M_i = \chi_{ij}H_j + \chi^{(2)}_{ijk}H_jH_k
,\end{align}
where $i,j,k \in \{x,y,z\}$ denote spatial directions, and repeated indices are summed over. The first term represents the conventional linear susceptibility, while the second term introduces a nonlinear, field-asymmetric contribution to the magnetization. The coefficient $\chi^{(2)}_{ijk}$ is a third-rank symmetric tensor that is even under spatial inversion and odd under time reversal, based on symmetry considerations. This parity and tensorial structure correspond to that of a magnetic octupole \cite{Hayami2018-rl, Hayami2018-zf, Hayami2024-vw}. Therefore, in materials with ferroic magnetic octupole order, such nonreciprocal magnetization responses with respect to field direction naturally emerge, providing a microscopic realization of the inverse single-sided magnet.

\begin{figure}[t!] 
\centering 
\includegraphics[width=0.45\textwidth]{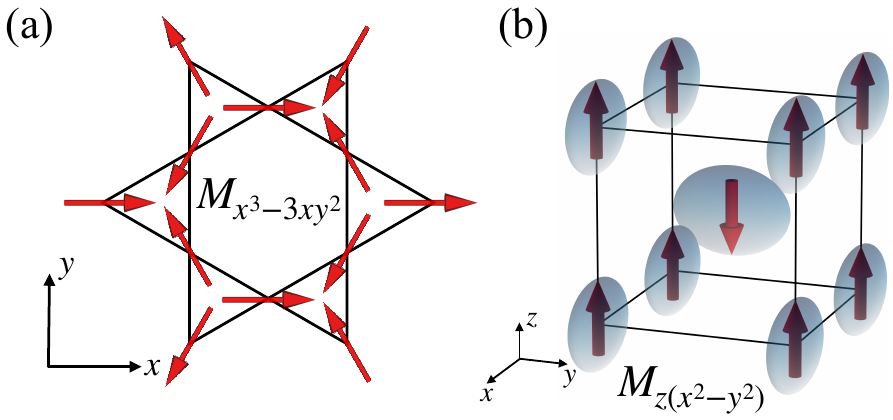} 
\caption{Examples of octupolar orders. (a) Coplanar model: a three-sublattice 120-degree magnetic structure on a kagom\'e lattice, also referred to as the all-in-all-out order. It represents $M_{x^3-3xy^2}$-type magnetic octupolar order. 
(b) Collinear model: N\'eel-type antiferromagnetic order in a crystal with antiferroic quadrupolar order. It exhibits $M_{z(x^2-y^2)}$-type magnetic octupolar order. Ellipsoids indicate the anisotropic charge distribution around magnetic ions, representing local quadrupolar environments. } 
\label{fig:Octupole} 
\end{figure}

{\it Model.---}To explore how inverse single-sided magnets can be realized in concrete systems, we examine two minimal models that host magnetic octupole order. The first, shown in Fig. \ref{fig:Octupole}(a), features a coplanar 120-degree magnetic structure on a kagom\'e lattice, also known as the all-in-all-out order. The second, illustrated in Fig. \ref{fig:Octupole}(b), represents a collinear antiferromagnet in a crystal with antiferroic electric quadrupole order. 
By analyzing these models, we demonstrate that the magnetization response becomes asymmetric with respect to the sign of the applied magnetic field. The coplanar model is potentially realizable in known magnetic materials, while the collinear model exhibits a diode-like response that sharply distinguishes between positive and negative field directions. 

We begin with the coplanar model, where the ground state at zero field forms a three-sublattice all-in-all-out structure. When an external magnetic field is applied along the $x$-direction, the spins cant within the plane. The effective classical Hamiltonian is given by
\begin{align}
	H_{\rm cop} &= J\sum_{i\neq j}\cos(\phi_i-\phi_j) - D\sum_{i}\cos \Bigl( 2\phi_i-\frac{4\pi i}{3} \Bigr) \nonumber \\
	&- H\sum_{i}\cos(\phi_i-\phi_H)
	\label{eq:H_cop}
,\end{align}
where $J > 0$ is the effective antiferromagnetic interaction, $D > 0$ is the in-plane single-ion anisotropy, $H$ is the magnitude of the applied magnetic field, and $\phi_H$ is its angle measured counterclockwise from the $x$-axis. The variables $\phi_i$ ($i \in \{1,2,3 \}$) denote the spin orientations on the three sublattices. A strong XY anisotropy is assumed, and the system can be treated as a two-dimensional classical spin model. We focus on the three-sublattice (${\bm q}={\bm 0}$) degrees of freedom. 
For $D<0$, the model remains equivalent under a 90-degree rotation of $\phi$'s. 

For a field along the $x$-axis ($\phi_H = 0$), assuming a smooth evolution from the zero-field state, the spin angles are taken as $\phi_3 = 0$, $\phi_1 = 2\pi/3 - \delta$, and $\phi_2 = 4\pi/3 + \delta$. Expanding the energy to the second order in $\delta$, we obtain
\begin{align} 
H_{\rm cop} = E_0 - \sqrt{3}H \delta + \frac{1}{2}( H_c - H )\delta^2 
,\end{align}
where $E_0 = -\tfrac{3}{2}J - 3D$ is the zero-field energy and $H_c = 3J + 8D$ is the characteristic field scale (not an exact critical field). The energy is minimized by
\begin{align}
\delta_{\rm min} = \frac{\sqrt{3}H}{H_c-H} \simeq \frac{\sqrt{3}H}{H_c} + \frac{\sqrt{3}H^2}{H_c^2} 
\label{eq:delta_min} 
,\end{align}
leading to the magnetization 
\begin{align}
M &= 1 + 2\cos \Bigl(\frac{2\pi}{3}-\delta_{\rm min} \Bigr) 
\simeq 3\biggl(\frac{H}{H_c}\biggr) + \frac{9}{2}\biggl(\frac{H}{H_c}\biggr)^2
\label{eq:M_cop}
.\end{align}
The nonreciprocal behavior is quantified by the ratio between the magnetization under positive and negative fields: 
\begin{align}
\frac{|M(H)|}{|M(-H)|} \simeq 1 + \frac{3H}{H_c} 
.\end{align}
This expression implies that a field equal to 10\% of $H_c$ yields a nonreciprocity of approximately 30\%. In terms of canting angle, this corresponds to tilt angles of about $11^\circ$ and $9^\circ$ for positive and negative fields, respectively.

\begin{figure}[t!] 
\centering 
\includegraphics[width=0.45\textwidth]{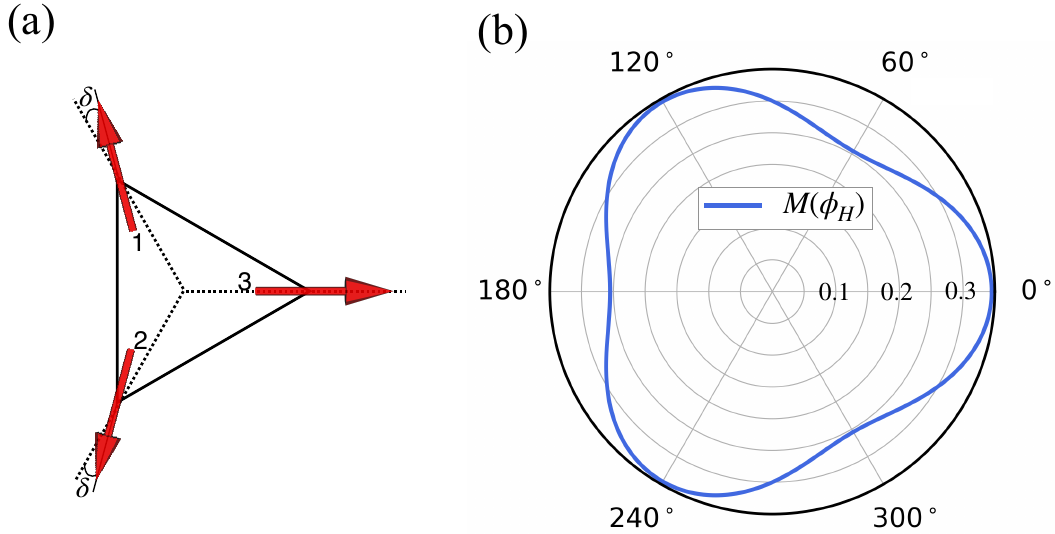} 
\caption{Coplanar model under external magnetic field. 
(a) Schematic illustration of spin canting when the magnetic field is applied along the $x$-axis. Sublattice spins $\phi_1$ and $\phi_2$ cant symmetrically toward the field direction.  
(b) Polar plot of the magnetization magnitude as a function of the field angle $\phi_H$ at $H/H_c = 0.1$.  
The radius indicates the magnetization, and the angle corresponds to the field direction. } 
\label{fig:Coplanar} 
\end{figure}

Finally, we consider the directional dependence of the magnetization. Up to the second order in $H$, symmetry constraints lead to
\begin{align}
	|M| \simeq 3\frac{|H|}{H_c} + \frac{9}{2}\biggl(\frac{H}{H_c}\biggr)^2\cos(3\phi_H)
.\end{align}
Figure \ref{fig:Coplanar}(b) shows the angular dependence at $H/H_c = 0.1$, demonstrating that $M \approx 0.35$ for fields along $+x$ and $M \approx 0.25$ for fields along $-x$, consistent with the expected nonreciprocal behavior. 

We now turn to the collinear model shown in Fig. \ref{fig:Octupole}(b), which represents a collinear antiferromagnet in a crystal with an antiferroic quadrupolar anisotropy. Such a configuration characterizes antiferromagnets with spin-split bands, also known as altermagnets \cite{Noda2016-tu, Naka2019-np, Hayami2019-qa, Hayami2020-bi, Smejkal2022-mr}. 
We consider systems composed of rare-earth or actinide ions, where strong spin-orbit coupling induces single-ion anisotropy. In particular, we assume that spins on sublattice 1 (at the corners of the rectangular prisms) rotate in the $zx$ plane, while those on sublattice 2 (at the centers) rotate in the $yz$ plane. 

The effective classical Hamiltonian reads
\begin{align}
	H_{\rm col} &= J\cos \phi_1 \cos \phi_2 - D\sum_{i=1,2}\cos(2\phi_i) \nonumber \\
	&- H\{ \cos(\phi_1-\phi_H) + \cos(\phi_2)\cos(\phi_H) \}
	\label{eq:H_col}
,\end{align}
where $J>0$ is the antiferromagnetic interaction, $D$ the single-ion anisotropy within the $zx$ ($yz$) plane for sublattice 1 (2), $H$ the magnetic field strength, and $\phi_H$ the field angle from the $z$-axis toward the $x$-axis. The spin angles $\phi_1$ and $\phi_2$ are measured from the $z$-axis to the $x$- and $y$-axis, respectively. We focus on the two-sublattice (${\bm q}={\bm 0}$) degrees of freedom. The zero-field ground state is assumed to be the N\'eel order along the $z$-axis, with sublattices 1 and 2 polarized along $+z$ and $-z$, respectively. The reversed domain is symmetrically equivalent under an inversion of the magnetic field. 

A key feature of this model is that the spin rotation planes differ between the two sublattices, unlike in conventional Ising or XY antiferromagnets. As a result, even when the single-ion anisotropy favors in-plane spin alignment ($D<0$), the system retains N\'eel order along the $z$-axis as long as $J+4D>0$. In contrast, Ising anisotropy ($D>0$) leads to conventional properties, such as spin-flip transition at $|H|=J$. However, the N\'eel-ordered state remains a local minimum of the energy for $|H|<J+4D$, as discussed later. 

\begin{figure}[t!] 
\centering 
\includegraphics[width=0.45\textwidth]{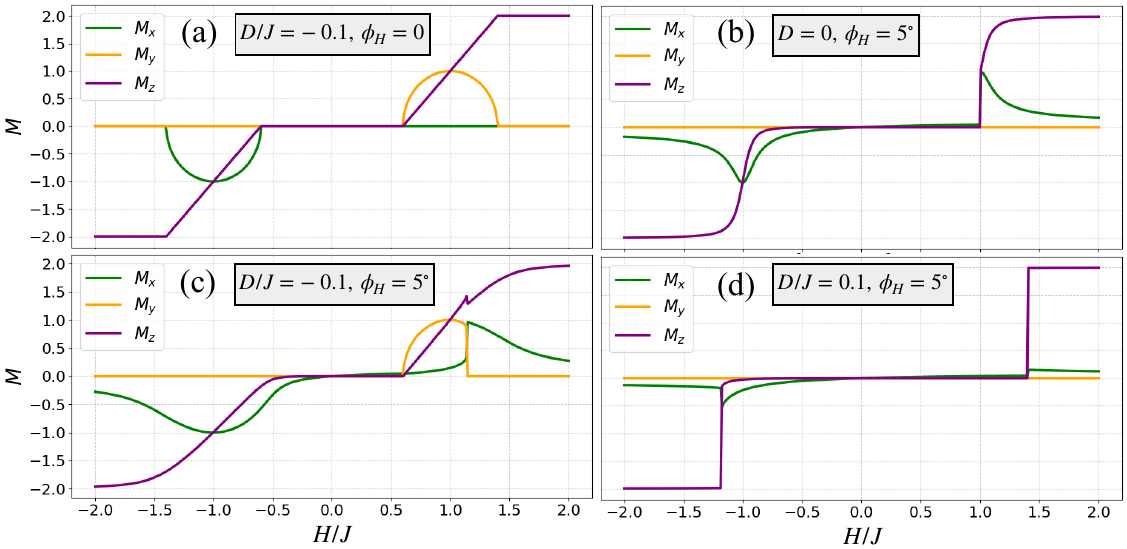} 
\caption{Field dependence of the magnetization in the collinear model for (a) $D=-0.1,\phi_H=0$, (b) $D=0,\phi_H=5^\circ$, (c) $D=-0.1,\phi_H=5^\circ$, and (d)$D=0.1,\phi_H=5^\circ$. Magnetization components $M_x$, $M_y$, and $M_z$ are shown in green, orange, and purple, respectively.} 
\label{fig:Collinear} 
\end{figure}

Let us now discuss the phase transition under a magnetic field applied along the positive $z$-axis for $D<0$. Substituting $\phi_H=\phi_1=0$ into Eq.~(\ref{eq:H_col}), we obtain
\begin{align} 
H_{\rm col} &= -D - H + (J - H)\cos \phi_2 - D\cos(2\phi_2) 
\label{eq:H_col_z}
.\end{align}
Minimizing this expression with respect to $\phi_2$ gives
\begin{align} 
\phi_2 = 
\begin{cases} \pi & (H < J + 4D) \\ 
\pm \cos^{-1} \left( \frac{-J + H}{4D} \right) & (J + 4D < H < J - 4D) \\ 
0 & (H > J - 4D) 
\end{cases} 
.\end{align}
This result indicates the emergence of a distinct intermediate-field phase. When the magnetic field is reversed, i.e., applied along the negative $z$-axis, the order parameter switches from $\phi_2$ to $\phi_1$. 
Notably, this ferromagnetic transition exhibits an unconventional form of nonreciprocity: the direction of the induced magnetization depends on the sign of the applied field. For positive $z$-fields, the $y$-axis component $M_y$ develops, whereas for negative $z$-fields, the $x$-axis component $M_x$ becomes dominant. 
This property arises from a coupling in the free energy of the form 
\begin{align}
\propto M_{z(x^2-y^2)}(M_x^2-M_y^2)H_z
,\end{align}
where $M_{z(x^2-y^2)}$ denotes the magnetic octupole.  Replacing $H_z$ with $M_z$ in this expression, i.e., 
\begin{align}
\propto M_{z(x^2-y^2)}(M_x^2-M_y^2)M_z
\end{align}
indicates that $M_z$ acts as a secondary order parameter coupled to the primary order parameter, $M_x$ or $M_y$, depending on the field direction. 
Figure~\ref{fig:Collinear}(a) shows the field dependence of the magnetization components for $D/J=-0.1$. Under positive fields, $M_y$ becomes the primary order parameter, while $M_x$ dominates under negative fields. In both cases, $M_z$ varies linearly with the field, consistent with its secondary nature. 

Next, we consider oblique fields in the $zx$ plane. For simplicity, we first focus on $D=0$. If no spin-flip occurs, Eq.~(\ref{eq:H_col}) simplifies to
\begin{align}
H_{\rm col} = H_z - \Delta \cos(\phi_1 - {\rm arg}[J+H_z+iH_x]) 
,\end{align}
where $(H_z, H_x) = H(\cos\phi_H, \sin\phi_H)$ and $\Delta=\sqrt{(J+H_z)^2+H_x^2}$. The energy is minimized when
\begin{align}
\phi_1 = {\rm arg}[J+H_z+iH_x]
\label{eq:phi_D_0}	
.\end{align}
As $H_z \to J$ and $H_x \to 0$, the canting angle approaches $\phi_1 \to 0$, while for $H_z \to -J$ with infinitesimal $H_x$, $\phi_1$ exhibits a $\pi/2$ rotation: $\phi_1 \to \text{sgn}(H_x)\pi/2$. This reveals strong nonreciprocity with respect to the field direction. 
Figure \ref{fig:Collinear}(b) presents numerical results for $\phi_H = 5^\circ$, which agree with the analytic result for $H<0$. For $H > 0$, a first-order transition to the polarized phase appears. This sharp contrast highlights the pronounced nonreciprocal response near $|H| \approx J$. 

We now examine the effect of finite anisotropy $D$. Figures \ref{fig:Collinear}(c,d) show the magnetization for $D/J = -0.1$ and $0.1$, respectively. For $D<0$, the second-order transitions become a crossover for negative fields, similar to the $D=0$ case. For positive fields, the transition from the N\'eel order to the ferromagnetic phase remains a second-order transition. In contrast, the transition to the polarized phase becomes first order since it involves the emergence of the $x$-component and vanishing of the $y$-component of the magnetization. The nonreciprocal nature is most pronounced near these transitions. 
For $D>0$, both signs of the field lead to first-order transitions but at different critical fields. For $H>0$, the spin-flip occurs near $H_c = J + 4D$, whereas for $H<0$, a tilted field induces spin-flipping at a lower threshold. For $D/J = 0.1$ and $\phi_H = 5^\circ$ in Fig. \ref{fig:Collinear}(d), the system is nearly unpolarized ($M \approx 0$) in the field range $1.2 \lessapprox |H|/J \lessapprox 1.4$ for $H > 0$, but fully polarized ($M \approx 2$) for $H < 0$, resulting in a giant nonreciprocal response. Although the N\'eel ordered phases for $|H|/J \gtrapprox 1$ are not the true ground state, they remain metastable, and the strong nonreciprocal behavior highlights the diode-like nature of inverse single-sided magnets.

{\it Discussions.---}We now discuss the physical properties, feasibility, and potential advances of the inverse single-sided magnet.  
From a scientific perspective, it is notable that an inversion of magnetic fields in octupolar magnets leads to symmetry-inequivalent states. This implies that the asymmetries observed in Figs. \ref{fig:Coplanar} and \ref{fig:Collinear} should manifest not only in magnetization but also in other physical properties such as electrical conductivity and spin transport. 
For instance, the anomalous Hall coefficient is expected to show a strong dependence on the direction of the applied field, similar to magnetization. In the collinear octupolar model, the ferromagnetic transition becomes a crossover only for one field direction. In this regime, magnetic susceptibility diverges only under the opposite field, indicating that nonreciprocal magnetic fluctuations with respect to field direction may strongly influence charge and spin transport \cite{Baibich1988-vn, Binasch1989-gf, Okamoto2019-sl, Okamoto2024-vp, Fang2023-xy}. 

The realization of an inverse single-sided magnet hinges on the stability of magnetic octupolar domains, particularly on the coercivity that prevents domain reversal. A given octupolar domain becomes metastable under fields of either sign. In order to maintain the nonreciprocal response, the coercivity must be large enough to suppress domain flipping. If domain inversion occurs, the magnetization exhibits a hysteresis loop analogous to the polarization–voltage hysteresis seen in antiferroelectrics, which can be linked to electric octupole order \cite{Dobal2001-hv}. 
In principle, magnetic octupolar domains can be aligned by applying a magnetic field with finite components $H_iH_jH_k$ ($i,j,k \in \{x,y,z\}$), which couple directly to the octupolar moment $M_{ijk}$. Field cooling in such a field is expected to align domains uniformly.
However, practical strategies for enhancing coercivity remain unclear. Nevertheless, since magnetic octupoles do not couple linearly to the applied field and domain inversion requires multiple spin flips, octupolar domains are expected to remain stable up to moderate field strengths.
In the collinear model, a field applied nearly along the $z$-axis couples weakly to the octupole, suggesting greater domain stability than larger tilted field. 

We now turn to potential material candidates for realizing an inverse single-sided magnet. Promising structures include those with magnetic ions occupying sites with ordered electric quadrupole moments, such as kagom\'e lattices and anisotropic tetragonal systems, as studied here.
Rare-earth and actinide ions are promising due to their strong spin-orbit coupling, leading to large single-ion anisotropy. Pseudo-kagom\'e 1-1-1 compounds such as HoAgGe \cite{Zhao2020-dl} and URhSn \cite{Shimizu2020-kb, Ishitobi2025-nn, Tabata_unpub, Tokunaga_unpub} exhibit 120-degree magnetic order, though the former does not form a three-sublattice structure. Furthermore, noncoplanar octupolar magnets include all-in-all-out ordered systems such as ${\rm Cd_2Os_2O_7}$ \cite{Yamaura2012-om, Shinaoka2012-uu} and rare-earth pyrochlore iridates $RE_2{\rm Ir_2O_7}$ ($RE =$ Nd, Sm, Eu, Tb) \cite{Tomiyasu2012-es, Sagayama2013-im, Lefrancois2015-uz, Donnerer2016-nx}. 
As for collinear models, we searched for compounds with rare-earth or actinide ions at quadrupolar sites in space groups $P4_2/mmc$, $P4_2/mcm$, and $P4_2/mnm$ in the MAGNDATA database \cite{Gallego2016-hq}, but found no reported examples. Nevertheless, the crystal structure database \cite{Grazulis2012-mb} lists rare-earth and actinide dioxides $RE{\rm O_2}$ with rutile structure (space group $P4_2/mcm$) as stable compounds. If such systems can be synthesized and exhibit N\'eel order, they could be candidates for inverse single-sided magnets. 

From an application perspective, inverse single-sided magnets function as diodes for magnetization, capable of suppressing magnetic interference. This makes them suitable for compact magnetic sensors, encoders, actuators, and related devices. When placed adjacent to ferromagnetic materials, they may also serve as directional magnetic shields, enabling microscopic single-sided magnets via single-sided shielding. As discussed above, not only magnetization but also other physical responses depend sensitively on field direction. This suggests the possibility of controlling responses to external stimuli, such as pressure, electric fields, or temperature gradients, via the sign of the magnetic field. Inverse single-sided magnets thus offer a platform for directionally selective magnetic functionalities at the microscale, with promising technological applications.

{\it Summary.---}In this study, we introduced the concept of an inverse single-sided magnet and demonstrated its realization in octupolar magnets. We analyzed two minimal models--a 120-degree-ordered system and a collinear-ordered system with site-dependent anisotropy--and found that both exhibit asymmetric magnetization responses with respect to the sign of the applied field. Notably, the collinear model displays strong polarization for one field direction while remaining nearly unresponsive for the opposite direction. Beyond magnetization, such asymmetry is expected to manifest in other physical quantities, offering fertile ground for future theoretical and experimental studies. Moreover, their diode-like behavior suggests promising applications in controlling magnetic interference, paving the way for advanced technological implementations. Like conventional single-sided magnets, they may serve as key building blocks in both scientific and engineering contexts.

{\it Acknowledgement.---}This work was supported by JSPS KAKENHI (Grant No. JP23K20824) from the Japan Society for the Promotion of Science.

\input{ISSM_ariv.bbl}

\end{document}

%% file: ISSM_ariv.bbl
%

%% file: ISSM_ariv.bbl
\begin{thebibliography}{36}%
\makeatletter
\providecommand \@ifxundefined [1]{%
 \@ifx{#1\undefined}
}%
\providecommand \@ifnum [1]{%
 \ifnum #1\expandafter \@firstoftwo
 \else \expandafter \@secondoftwo
 \fi
}%
\providecommand \@ifx [1]{%
 \ifx #1\expandafter \@firstoftwo
 \else \expandafter \@secondoftwo
 \fi
}%
\providecommand \natexlab [1]{#1}%
\providecommand \enquote  [1]{``#1''}%
\providecommand \bibnamefont  [1]{#1}%
\providecommand \bibfnamefont [1]{#1}%
\providecommand \citenamefont [1]{#1}%
\providecommand \href@noop [0]{\@secondoftwo}%
\providecommand \href [0]{\begingroup \@sanitize@url \@href}%
\providecommand \@href[1]{\@@startlink{#1}\@@href}%
\providecommand \@@href[1]{\endgroup#1\@@endlink}%
\providecommand \@sanitize@url [0]{\catcode `\\12\catcode `\$12\catcode
  `\&12\catcode `\#12\catcode `\^12\catcode `\_12\catcode `\%12\relax}%
\providecommand \@@startlink[1]{}%
\providecommand \@@endlink[0]{}%
\providecommand \url  [0]{\begingroup\@sanitize@url \@url }%
\providecommand \@url [1]{\endgroup\@href {#1}{\urlprefix }}%
\providecommand \urlprefix  [0]{URL }%
\providecommand \Eprint [0]{\href }%
\providecommand \doibase [0]{https://doi.org/}%
\providecommand \selectlanguage [0]{\@gobble}%
\providecommand \bibinfo  [0]{\@secondoftwo}%
\providecommand \bibfield  [0]{\@secondoftwo}%
\providecommand \translation [1]{[#1]}%
\providecommand \BibitemOpen [0]{}%
\providecommand \bibitemStop [0]{}%
\providecommand \bibitemNoStop [0]{.\EOS\space}%
\providecommand \EOS [0]{\spacefactor3000\relax}%
\providecommand \BibitemShut  [1]{\csname bibitem#1\endcsname}%
\let\auto@bib@innerbib\@empty
\bibitem [{\citenamefont {Mallinson}(1973)}]{Mallinson1973-xc}%
  \BibitemOpen
  \bibfield  {author} {\bibinfo {author} {\bibfnamefont {J.}~\bibnamefont
  {Mallinson}},\ }\href {https://doi.org/10.1109/tmag.1973.1067714} {\bibfield
  {journal} {\bibinfo  {journal} {IEEE Trans. Magn.}\ }\textbf {\bibinfo
  {volume} {9}},\ \bibinfo {pages} {678} (\bibinfo {year} {1973})}\BibitemShut
  {NoStop}%
\bibitem [{\citenamefont {Halbach}(1979)}]{Halbach1979-fb}%
  \BibitemOpen
  \bibfield  {author} {\bibinfo {author} {\bibfnamefont {K.}~\bibnamefont
  {Halbach}},\ }\href {https://doi.org/10.1109/tns.1979.4330638} {\bibfield
  {journal} {\bibinfo  {journal} {IEEE Trans. Nucl. Sci.}\ }\textbf {\bibinfo
  {volume} {26}},\ \bibinfo {pages} {3882} (\bibinfo {year}
  {1979})}\BibitemShut {NoStop}%
\bibitem [{\citenamefont {Halbach}(1980)}]{Halbach1980-gi}%
  \BibitemOpen
  \bibfield  {author} {\bibinfo {author} {\bibfnamefont {K.}~\bibnamefont
  {Halbach}},\ }\href {https://doi.org/10.1016/0029-554x(80)90094-4} {\bibfield
   {journal} {\bibinfo  {journal} {Nucl. Instrum. Meth.}\ }\textbf {\bibinfo
  {volume} {169}},\ \bibinfo {pages} {1} (\bibinfo {year} {1980})}\BibitemShut
  {NoStop}%
\bibitem [{\citenamefont {Halbach}(1985)}]{Halbach1985-sd}%
  \BibitemOpen
  \bibfield  {author} {\bibinfo {author} {\bibfnamefont {K.}~\bibnamefont
  {Halbach}},\ }\href {https://doi.org/10.1063/1.335021} {\bibfield  {journal}
  {\bibinfo  {journal} {J. Appl. Phys.}\ }\textbf {\bibinfo {volume} {57}},\
  \bibinfo {pages} {3605} (\bibinfo {year} {1985})}\BibitemShut {NoStop}%
\bibitem [{\citenamefont {Zhu}\ and\ \citenamefont {Howe}(2001)}]{Zhu2001-ju}%
  \BibitemOpen
  \bibfield  {author} {\bibinfo {author} {\bibfnamefont {Z.~Q.}\ \bibnamefont
  {Zhu}}\ and\ \bibinfo {author} {\bibfnamefont {D.}~\bibnamefont {Howe}},\
  }\href {https://doi.org/10.1049/ip-epa:20010479} {\bibfield  {journal}
  {\bibinfo  {journal} {IEE Proc. - Electr. Power Appl.}\ }\textbf {\bibinfo
  {volume} {148}},\ \bibinfo {pages} {299} (\bibinfo {year}
  {2001})}\BibitemShut {NoStop}%
\bibitem [{\citenamefont {Coey}(2002)}]{Coey2002-ht}%
  \BibitemOpen
  \bibfield  {author} {\bibinfo {author} {\bibfnamefont {J.~M.~D.}\
  \bibnamefont {Coey}},\ }\href {https://doi.org/10.1016/s0304-8853(02)00335-9}
  {\bibfield  {journal} {\bibinfo  {journal} {J. Magn. Magn. Mater.}\ }\textbf
  {\bibinfo {volume} {248}},\ \bibinfo {pages} {441} (\bibinfo {year}
  {2002})}\BibitemShut {NoStop}%
\bibitem [{\citenamefont {Gieras}\ \emph {et~al.}()\citenamefont {Gieras},
  \citenamefont {Wang},\ and\ \citenamefont {Kamper}}]{Gieras2008-zh}%
  \BibitemOpen
  \bibfield  {author} {\bibinfo {author} {\bibfnamefont {J.~F.}\ \bibnamefont
  {Gieras}}, \bibinfo {author} {\bibfnamefont {R.-J.}\ \bibnamefont {Wang}},\
  and\ \bibinfo {author} {\bibfnamefont {M.~J.}\ \bibnamefont {Kamper}},\
  }\href@noop {} {\bibinfo {title}
  {\href{https://link.springer.com/book/10.1007/978-1-4020-8227-6}{Axial Flux
  Permanent Magnet Brushless Machines}}},\ \bibinfo {note} {(Springer, New
  York, NY, 2008).}\BibitemShut {Stop}%
\bibitem [{Gie()}]{Gieras2009-zb}%
  \BibitemOpen
  \href@noop {} {}\bibinfo {note} {J. F. Gieras,
  \href{https://api.taylorfrancis.com/content/books/mono/download?identifierName=doi&identifierValue=10.1201/9780429292736&type=googlepdf}{Permanent
  Magnet Motor Technology: Design and applications}, (Taylor \& Francis,
  London, England, 2009).}\BibitemShut {Stop}%
\bibitem [{Spa()}]{Spaldin2010-qg}%
  \BibitemOpen
  \href@noop {} {}\bibinfo {note} {N. A. Spaldin,
  \href{https://www.cambridge.org/jp/universitypress/subjects/engineering/materials-science/magnetic-materials-fundamentals-and-applications-2nd-edition?format=HB&isbn=9780521886697}{Magnetic
  materials: Fundamentals and applications}, (Cambridge University Press,
  Cambridge, England, 2010).}\BibitemShut {Stop}%
\bibitem [{\citenamefont {Hayami}\ and\ \citenamefont
  {Kusunose}(2018)}]{Hayami2018-rl}%
  \BibitemOpen
  \bibfield  {author} {\bibinfo {author} {\bibfnamefont {S.}~\bibnamefont
  {Hayami}}\ and\ \bibinfo {author} {\bibfnamefont {H.}~\bibnamefont
  {Kusunose}},\ }\href {https://doi.org/10.7566/JPSJ.87.033709} {\bibfield
  {journal} {\bibinfo  {journal} {J. Phys. Soc. Jpn.}\ }\textbf {\bibinfo
  {volume} {87}},\ \bibinfo {pages} {1} (\bibinfo {year} {2018})}\BibitemShut
  {NoStop}%
\bibitem [{\citenamefont {Hayami}\ \emph {et~al.}(2018)\citenamefont {Hayami},
  \citenamefont {Yatsushiro}, \citenamefont {Yanagi},\ and\ \citenamefont
  {Kusunose}}]{Hayami2018-zf}%
  \BibitemOpen
  \bibfield  {author} {\bibinfo {author} {\bibfnamefont {S.}~\bibnamefont
  {Hayami}}, \bibinfo {author} {\bibfnamefont {M.}~\bibnamefont {Yatsushiro}},
  \bibinfo {author} {\bibfnamefont {Y.}~\bibnamefont {Yanagi}},\ and\ \bibinfo
  {author} {\bibfnamefont {H.}~\bibnamefont {Kusunose}},\ }\href
  {https://doi.org/10.1103/PhysRevB.98.165110} {\bibfield  {journal} {\bibinfo
  {journal} {Phys. Rev. B}\ }\textbf {\bibinfo {volume} {98}},\ \bibinfo
  {pages} {165110} (\bibinfo {year} {2018})}\BibitemShut {NoStop}%
\bibitem [{\citenamefont {Hayami}\ and\ \citenamefont
  {Kusunose}(2024)}]{Hayami2024-vw}%
  \BibitemOpen
  \bibfield  {author} {\bibinfo {author} {\bibfnamefont {S.}~\bibnamefont
  {Hayami}}\ and\ \bibinfo {author} {\bibfnamefont {H.}~\bibnamefont
  {Kusunose}},\ }\href {https://doi.org/10.7566/JPSJ.93.072001} {\bibfield
  {journal} {\bibinfo  {journal} {J. Phys. Soc. Jpn.}\ }\textbf {\bibinfo
  {volume} {93}},\ \bibinfo {pages} {072001} (\bibinfo {year}
  {2024})}\BibitemShut {NoStop}%
\bibitem [{\citenamefont {Noda}\ \emph {et~al.}(2016)\citenamefont {Noda},
  \citenamefont {Ohno},\ and\ \citenamefont {Nakamura}}]{Noda2016-tu}%
  \BibitemOpen
  \bibfield  {author} {\bibinfo {author} {\bibfnamefont {Y.}~\bibnamefont
  {Noda}}, \bibinfo {author} {\bibfnamefont {K.}~\bibnamefont {Ohno}},\ and\
  \bibinfo {author} {\bibfnamefont {S.}~\bibnamefont {Nakamura}},\ }\href
  {https://doi.org/10.1039/c5cp07806g} {\bibfield  {journal} {\bibinfo
  {journal} {Phys. Chem. Chem. Phys.}\ }\textbf {\bibinfo {volume} {18}},\
  \bibinfo {pages} {13294} (\bibinfo {year} {2016})}\BibitemShut {NoStop}%
\bibitem [{\citenamefont {Naka}\ \emph {et~al.}(2019)\citenamefont {Naka},
  \citenamefont {Hayami}, \citenamefont {Kusunose}, \citenamefont {Yanagi},
  \citenamefont {Motome},\ and\ \citenamefont {Seo}}]{Naka2019-np}%
  \BibitemOpen
  \bibfield  {author} {\bibinfo {author} {\bibfnamefont {M.}~\bibnamefont
  {Naka}}, \bibinfo {author} {\bibfnamefont {S.}~\bibnamefont {Hayami}},
  \bibinfo {author} {\bibfnamefont {H.}~\bibnamefont {Kusunose}}, \bibinfo
  {author} {\bibfnamefont {Y.}~\bibnamefont {Yanagi}}, \bibinfo {author}
  {\bibfnamefont {Y.}~\bibnamefont {Motome}},\ and\ \bibinfo {author}
  {\bibfnamefont {H.}~\bibnamefont {Seo}},\ }\href
  {https://doi.org/10.1038/s41467-019-12229-y} {\bibfield  {journal} {\bibinfo
  {journal} {Nat. Commun.}\ }\textbf {\bibinfo {volume} {10}},\ \bibinfo
  {pages} {4305} (\bibinfo {year} {2019})}\BibitemShut {NoStop}%
\bibitem [{\citenamefont {Hayami}\ \emph {et~al.}(2019)\citenamefont {Hayami},
  \citenamefont {Yanagi},\ and\ \citenamefont {Kusunose}}]{Hayami2019-qa}%
  \BibitemOpen
  \bibfield  {author} {\bibinfo {author} {\bibfnamefont {S.}~\bibnamefont
  {Hayami}}, \bibinfo {author} {\bibfnamefont {Y.}~\bibnamefont {Yanagi}},\
  and\ \bibinfo {author} {\bibfnamefont {H.}~\bibnamefont {Kusunose}},\ }\href
  {https://doi.org/10.7566/JPSJ.88.123702} {\bibfield  {journal} {\bibinfo
  {journal} {J. Phys. Soc. Jpn.}\ }\textbf {\bibinfo {volume} {88}},\ \bibinfo
  {pages} {123702} (\bibinfo {year} {2019})}\BibitemShut {NoStop}%
\bibitem [{\citenamefont {Hayami}\ \emph {et~al.}(2020)\citenamefont {Hayami},
  \citenamefont {Yanagi},\ and\ \citenamefont {Kusunose}}]{Hayami2020-bi}%
  \BibitemOpen
  \bibfield  {author} {\bibinfo {author} {\bibfnamefont {S.}~\bibnamefont
  {Hayami}}, \bibinfo {author} {\bibfnamefont {Y.}~\bibnamefont {Yanagi}},\
  and\ \bibinfo {author} {\bibfnamefont {H.}~\bibnamefont {Kusunose}},\ }\href
  {https://doi.org/10.1103/PhysRevB.102.144441} {\bibfield  {journal} {\bibinfo
   {journal} {Phys. Rev. B}\ }\textbf {\bibinfo {volume} {102}},\ \bibinfo
  {pages} {144441} (\bibinfo {year} {2020})}\BibitemShut {NoStop}%
\bibitem [{\citenamefont {Šmejkal}\ \emph {et~al.}(2022)\citenamefont
  {Šmejkal}, \citenamefont {Sinova},\ and\ \citenamefont
  {Jungwirth}}]{Smejkal2022-mr}%
  \BibitemOpen
  \bibfield  {author} {\bibinfo {author} {\bibfnamefont {L.}~\bibnamefont
  {Šmejkal}}, \bibinfo {author} {\bibfnamefont {J.}~\bibnamefont {Sinova}},\
  and\ \bibinfo {author} {\bibfnamefont {T.}~\bibnamefont {Jungwirth}},\ }\href
  {https://doi.org/10.1103/PhysRevX.12.040501} {\bibfield  {journal} {\bibinfo
  {journal} {Phys. Rev. X}\ }\textbf {\bibinfo {volume} {12}},\ \bibinfo
  {pages} {040501} (\bibinfo {year} {2022})}\BibitemShut {NoStop}%
\bibitem [{\citenamefont {Baibich}\ \emph {et~al.}(1988)\citenamefont
  {Baibich}, \citenamefont {Broto}, \citenamefont {Fert}, \citenamefont
  {{Nguyen Van Dau F}}, \citenamefont {Petroff}, \citenamefont {Etienne},
  \citenamefont {Creuzet}, \citenamefont {Friederich},\ and\ \citenamefont
  {Chazelas}}]{Baibich1988-vn}%
  \BibitemOpen
  \bibfield  {author} {\bibinfo {author} {\bibfnamefont {M.~N.}\ \bibnamefont
  {Baibich}}, \bibinfo {author} {\bibfnamefont {J.~M.}\ \bibnamefont {Broto}},
  \bibinfo {author} {\bibfnamefont {A.}~\bibnamefont {Fert}}, \bibinfo {author}
  {\bibnamefont {{Nguyen Van Dau F}}}, \bibinfo {author} {\bibfnamefont
  {F.}~\bibnamefont {Petroff}}, \bibinfo {author} {\bibfnamefont
  {P.}~\bibnamefont {Etienne}}, \bibinfo {author} {\bibfnamefont
  {G.}~\bibnamefont {Creuzet}}, \bibinfo {author} {\bibfnamefont
  {A.}~\bibnamefont {Friederich}},\ and\ \bibinfo {author} {\bibfnamefont
  {J.}~\bibnamefont {Chazelas}},\ }\href
  {https://doi.org/10.1103/PhysRevLett.61.2472} {\bibfield  {journal} {\bibinfo
   {journal} {Phys. Rev. Lett.}\ }\textbf {\bibinfo {volume} {61}},\ \bibinfo
  {pages} {2472} (\bibinfo {year} {1988})}\BibitemShut {NoStop}%
\bibitem [{\citenamefont {Binasch}\ \emph {et~al.}(1989)\citenamefont
  {Binasch}, \citenamefont {Grünberg}, \citenamefont {Saurenbach},\ and\
  \citenamefont {Zinn}}]{Binasch1989-gf}%
  \BibitemOpen
  \bibfield  {author} {\bibinfo {author} {\bibfnamefont {G.}~\bibnamefont
  {Binasch}}, \bibinfo {author} {\bibfnamefont {P.}~\bibnamefont {Grünberg}},
  \bibinfo {author} {\bibfnamefont {F.}~\bibnamefont {Saurenbach}},\ and\
  \bibinfo {author} {\bibfnamefont {W.}~\bibnamefont {Zinn}},\ }\href
  {https://doi.org/10.1103/physrevb.39.4828} {\bibfield  {journal} {\bibinfo
  {journal} {Phys. Rev. B}\ }\textbf {\bibinfo {volume} {39}},\ \bibinfo
  {pages} {4828} (\bibinfo {year} {1989})}\BibitemShut {NoStop}%
\bibitem [{\citenamefont {Okamoto}\ \emph {et~al.}(2019)\citenamefont
  {Okamoto}, \citenamefont {Egami},\ and\ \citenamefont
  {Nagaosa}}]{Okamoto2019-sl}%
  \BibitemOpen
  \bibfield  {author} {\bibinfo {author} {\bibfnamefont {S.}~\bibnamefont
  {Okamoto}}, \bibinfo {author} {\bibfnamefont {T.}~\bibnamefont {Egami}},\
  and\ \bibinfo {author} {\bibfnamefont {N.}~\bibnamefont {Nagaosa}},\ }\href
  {https://doi.org/10.1103/physrevlett.123.196603} {\bibfield  {journal}
  {\bibinfo  {journal} {Phys. Rev. Lett.}\ }\textbf {\bibinfo {volume} {123}},\
  \bibinfo {pages} {196603} (\bibinfo {year} {2019})}\BibitemShut {NoStop}%
\bibitem [{\citenamefont {Okamoto}\ and\ \citenamefont
  {Nagaosa}(2024)}]{Okamoto2024-vp}%
  \BibitemOpen
  \bibfield  {author} {\bibinfo {author} {\bibfnamefont {S.}~\bibnamefont
  {Okamoto}}\ and\ \bibinfo {author} {\bibfnamefont {N.}~\bibnamefont
  {Nagaosa}},\ }\href {https://doi.org/10.1038/s41535-024-00631-9} {\bibfield
  {journal} {\bibinfo  {journal} {Npj Quantum Mater.}\ }\textbf {\bibinfo
  {volume} {9}},\ \bibinfo {pages} {1} (\bibinfo {year} {2024})}\BibitemShut
  {NoStop}%
\bibitem [{\citenamefont {Fang}\ \emph {et~al.}(2023)\citenamefont {Fang},
  \citenamefont {Wan}, \citenamefont {Zhang}, \citenamefont {Okamoto},
  \citenamefont {Ma}, \citenamefont {Qin}, \citenamefont {Wang}, \citenamefont
  {Guo}, \citenamefont {Dong}, \citenamefont {Yu}, \citenamefont {Wen},
  \citenamefont {Tang}, \citenamefont {Parkin}, \citenamefont {Nagaosa},
  \citenamefont {Lu},\ and\ \citenamefont {Han}}]{Fang2023-xy}%
  \BibitemOpen
  \bibfield  {author} {\bibinfo {author} {\bibfnamefont {C.}~\bibnamefont
  {Fang}}, \bibinfo {author} {\bibfnamefont {C.}~\bibnamefont {Wan}}, \bibinfo
  {author} {\bibfnamefont {X.}~\bibnamefont {Zhang}}, \bibinfo {author}
  {\bibfnamefont {S.}~\bibnamefont {Okamoto}}, \bibinfo {author} {\bibfnamefont
  {T.}~\bibnamefont {Ma}}, \bibinfo {author} {\bibfnamefont {J.}~\bibnamefont
  {Qin}}, \bibinfo {author} {\bibfnamefont {X.}~\bibnamefont {Wang}}, \bibinfo
  {author} {\bibfnamefont {C.}~\bibnamefont {Guo}}, \bibinfo {author}
  {\bibfnamefont {J.}~\bibnamefont {Dong}}, \bibinfo {author} {\bibfnamefont
  {G.}~\bibnamefont {Yu}}, \bibinfo {author} {\bibfnamefont {Z.}~\bibnamefont
  {Wen}}, \bibinfo {author} {\bibfnamefont {N.}~\bibnamefont {Tang}}, \bibinfo
  {author} {\bibfnamefont {S.~S.~P.}\ \bibnamefont {Parkin}}, \bibinfo {author}
  {\bibfnamefont {N.}~\bibnamefont {Nagaosa}}, \bibinfo {author} {\bibfnamefont
  {Y.}~\bibnamefont {Lu}},\ and\ \bibinfo {author} {\bibfnamefont
  {X.}~\bibnamefont {Han}},\ }\href
  {https://doi.org/10.1021/acs.nanolett.3c03085} {\bibfield  {journal}
  {\bibinfo  {journal} {Nano Lett.}\ }\textbf {\bibinfo {volume} {23}},\
  \bibinfo {pages} {11485} (\bibinfo {year} {2023})}\BibitemShut {NoStop}%
\bibitem [{\citenamefont {Dobal}\ \emph {et~al.}(2001)\citenamefont {Dobal},
  \citenamefont {Katiyar}, \citenamefont {Bharadwaja},\ and\ \citenamefont
  {Krupanidhi}}]{Dobal2001-hv}%
  \BibitemOpen
  \bibfield  {author} {\bibinfo {author} {\bibfnamefont {P.~S.}\ \bibnamefont
  {Dobal}}, \bibinfo {author} {\bibfnamefont {R.~S.}\ \bibnamefont {Katiyar}},
  \bibinfo {author} {\bibfnamefont {S.~S.~N.}\ \bibnamefont {Bharadwaja}},\
  and\ \bibinfo {author} {\bibfnamefont {S.~B.}\ \bibnamefont {Krupanidhi}},\
  }\href {https://doi.org/10.1063/1.1356730} {\bibfield  {journal} {\bibinfo
  {journal} {Appl. Phys. Lett.}\ }\textbf {\bibinfo {volume} {78}},\ \bibinfo
  {pages} {1730} (\bibinfo {year} {2001})}\BibitemShut {NoStop}%
\bibitem [{\citenamefont {Zhao}\ \emph {et~al.}(2020)\citenamefont {Zhao},
  \citenamefont {Deng}, \citenamefont {Chen}, \citenamefont {Ross},
  \citenamefont {Petříček}, \citenamefont {Günther}, \citenamefont
  {Russina}, \citenamefont {Hutanu},\ and\ \citenamefont
  {Gegenwart}}]{Zhao2020-dl}%
  \BibitemOpen
  \bibfield  {author} {\bibinfo {author} {\bibfnamefont {K.}~\bibnamefont
  {Zhao}}, \bibinfo {author} {\bibfnamefont {H.}~\bibnamefont {Deng}}, \bibinfo
  {author} {\bibfnamefont {H.}~\bibnamefont {Chen}}, \bibinfo {author}
  {\bibfnamefont {K.~A.}\ \bibnamefont {Ross}}, \bibinfo {author}
  {\bibfnamefont {V.}~\bibnamefont {Petříček}}, \bibinfo {author}
  {\bibfnamefont {G.}~\bibnamefont {Günther}}, \bibinfo {author}
  {\bibfnamefont {M.}~\bibnamefont {Russina}}, \bibinfo {author} {\bibfnamefont
  {V.}~\bibnamefont {Hutanu}},\ and\ \bibinfo {author} {\bibfnamefont
  {P.}~\bibnamefont {Gegenwart}},\ }\href
  {https://doi.org/10.1126/science.aaw1666} {\bibfield  {journal} {\bibinfo
  {journal} {Science}\ }\textbf {\bibinfo {volume} {367}},\ \bibinfo {pages}
  {1218} (\bibinfo {year} {2020})}\BibitemShut {NoStop}%
\bibitem [{\citenamefont {Shimizu}\ \emph {et~al.}(2020)\citenamefont
  {Shimizu}, \citenamefont {Miyake}, \citenamefont {Maurya}, \citenamefont
  {Honda}, \citenamefont {Nakamura}, \citenamefont {Sato}, \citenamefont {Li},
  \citenamefont {Homma}, \citenamefont {Yokoyama}, \citenamefont {Tokunaga},
  \citenamefont {Tokunaga},\ and\ \citenamefont {Aoki}}]{Shimizu2020-kb}%
  \BibitemOpen
  \bibfield  {author} {\bibinfo {author} {\bibfnamefont {Y.}~\bibnamefont
  {Shimizu}}, \bibinfo {author} {\bibfnamefont {A.}~\bibnamefont {Miyake}},
  \bibinfo {author} {\bibfnamefont {A.}~\bibnamefont {Maurya}}, \bibinfo
  {author} {\bibfnamefont {F.}~\bibnamefont {Honda}}, \bibinfo {author}
  {\bibfnamefont {A.}~\bibnamefont {Nakamura}}, \bibinfo {author}
  {\bibfnamefont {Y.~J.}\ \bibnamefont {Sato}}, \bibinfo {author}
  {\bibfnamefont {D.}~\bibnamefont {Li}}, \bibinfo {author} {\bibfnamefont
  {Y.}~\bibnamefont {Homma}}, \bibinfo {author} {\bibfnamefont
  {M.}~\bibnamefont {Yokoyama}}, \bibinfo {author} {\bibfnamefont
  {Y.}~\bibnamefont {Tokunaga}}, \bibinfo {author} {\bibfnamefont
  {M.}~\bibnamefont {Tokunaga}},\ and\ \bibinfo {author} {\bibfnamefont
  {D.}~\bibnamefont {Aoki}},\ }\href
  {https://doi.org/10.1103/PhysRevB.102.134411} {\bibfield  {journal} {\bibinfo
   {journal} {Phys. Rev. B}\ }\textbf {\bibinfo {volume} {102}},\ \bibinfo
  {pages} {134411} (\bibinfo {year} {2020})}\BibitemShut {NoStop}%
\bibitem [{Ish()}]{Ishitobi2025-nn}%
  \BibitemOpen
  \href@noop {} {}\bibinfo {note} {T. Ishitobi and K. Hattori,
  \href{http://arxiv.org/abs/2502.13977}{arxiv:2502.13977}.}\BibitemShut
  {Stop}%
\bibitem [{Tab()}]{Tabata_unpub}%
  \BibitemOpen
  \href@noop {} {}\bibinfo {note} {C. Tabata {\it et. al.},
  unpublished}\BibitemShut {NoStop}%
\bibitem [{Tok()}]{Tokunaga_unpub}%
  \BibitemOpen
  \href@noop {} {}\bibinfo {note} {Y. Tokunaga {\it et. al.},
  unpublished}\BibitemShut {NoStop}%
\bibitem [{\citenamefont {Yamaura}\ \emph {et~al.}(2012)\citenamefont
  {Yamaura}, \citenamefont {Ohgushi}, \citenamefont {Ohsumi}, \citenamefont
  {Hasegawa}, \citenamefont {Yamauchi}, \citenamefont {Sugimoto}, \citenamefont
  {Takeshita}, \citenamefont {Tokuda}, \citenamefont {Takata}, \citenamefont
  {Udagawa}, \citenamefont {Takigawa}, \citenamefont {Harima}, \citenamefont
  {Arima},\ and\ \citenamefont {Hiroi}}]{Yamaura2012-om}%
  \BibitemOpen
  \bibfield  {author} {\bibinfo {author} {\bibfnamefont {J.}~\bibnamefont
  {Yamaura}}, \bibinfo {author} {\bibfnamefont {K.}~\bibnamefont {Ohgushi}},
  \bibinfo {author} {\bibfnamefont {H.}~\bibnamefont {Ohsumi}}, \bibinfo
  {author} {\bibfnamefont {T.}~\bibnamefont {Hasegawa}}, \bibinfo {author}
  {\bibfnamefont {I.}~\bibnamefont {Yamauchi}}, \bibinfo {author}
  {\bibfnamefont {K.}~\bibnamefont {Sugimoto}}, \bibinfo {author}
  {\bibfnamefont {S.}~\bibnamefont {Takeshita}}, \bibinfo {author}
  {\bibfnamefont {A.}~\bibnamefont {Tokuda}}, \bibinfo {author} {\bibfnamefont
  {M.}~\bibnamefont {Takata}}, \bibinfo {author} {\bibfnamefont
  {M.}~\bibnamefont {Udagawa}}, \bibinfo {author} {\bibfnamefont
  {M.}~\bibnamefont {Takigawa}}, \bibinfo {author} {\bibfnamefont
  {H.}~\bibnamefont {Harima}}, \bibinfo {author} {\bibfnamefont
  {T.}~\bibnamefont {Arima}},\ and\ \bibinfo {author} {\bibfnamefont
  {Z.}~\bibnamefont {Hiroi}},\ }\href
  {https://doi.org/10.1103/PhysRevLett.108.247205} {\bibfield  {journal}
  {\bibinfo  {journal} {Phys. Rev. Lett.}\ }\textbf {\bibinfo {volume} {108}},\
  \bibinfo {pages} {247205} (\bibinfo {year} {2012})}\BibitemShut {NoStop}%
\bibitem [{\citenamefont {Shinaoka}\ \emph {et~al.}(2012)\citenamefont
  {Shinaoka}, \citenamefont {Miyake},\ and\ \citenamefont
  {Ishibashi}}]{Shinaoka2012-uu}%
  \BibitemOpen
  \bibfield  {author} {\bibinfo {author} {\bibfnamefont {H.}~\bibnamefont
  {Shinaoka}}, \bibinfo {author} {\bibfnamefont {T.}~\bibnamefont {Miyake}},\
  and\ \bibinfo {author} {\bibfnamefont {S.}~\bibnamefont {Ishibashi}},\ }\href
  {https://doi.org/10.1103/PhysRevLett.108.247204} {\bibfield  {journal}
  {\bibinfo  {journal} {Phys. Rev. Lett.}\ }\textbf {\bibinfo {volume} {108}},\
  \bibinfo {pages} {247204} (\bibinfo {year} {2012})}\BibitemShut {NoStop}%
\bibitem [{\citenamefont {Tomiyasu}\ \emph {et~al.}(2012)\citenamefont
  {Tomiyasu}, \citenamefont {Matsuhira}, \citenamefont {Iwasa}, \citenamefont
  {Watahiki}, \citenamefont {Takagi}, \citenamefont {Wakeshima}, \citenamefont
  {Hinatsu}, \citenamefont {Yokoyama}, \citenamefont {Ohoyama},\ and\
  \citenamefont {Yamada}}]{Tomiyasu2012-es}%
  \BibitemOpen
  \bibfield  {author} {\bibinfo {author} {\bibfnamefont {K.}~\bibnamefont
  {Tomiyasu}}, \bibinfo {author} {\bibfnamefont {K.}~\bibnamefont {Matsuhira}},
  \bibinfo {author} {\bibfnamefont {K.}~\bibnamefont {Iwasa}}, \bibinfo
  {author} {\bibfnamefont {M.}~\bibnamefont {Watahiki}}, \bibinfo {author}
  {\bibfnamefont {S.}~\bibnamefont {Takagi}}, \bibinfo {author} {\bibfnamefont
  {M.}~\bibnamefont {Wakeshima}}, \bibinfo {author} {\bibfnamefont
  {Y.}~\bibnamefont {Hinatsu}}, \bibinfo {author} {\bibfnamefont
  {M.}~\bibnamefont {Yokoyama}}, \bibinfo {author} {\bibfnamefont
  {K.}~\bibnamefont {Ohoyama}},\ and\ \bibinfo {author} {\bibfnamefont
  {K.}~\bibnamefont {Yamada}},\ }\href {https://doi.org/10.1143/JPSJ.81.034709}
  {\bibfield  {journal} {\bibinfo  {journal} {J. Phys. Soc. Jpn.}\ }\textbf
  {\bibinfo {volume} {81}},\ \bibinfo {pages} {034709} (\bibinfo {year}
  {2012})}\BibitemShut {NoStop}%
\bibitem [{\citenamefont {Sagayama}\ \emph {et~al.}(2013)\citenamefont
  {Sagayama}, \citenamefont {Uematsu}, \citenamefont {Arima}, \citenamefont
  {Sugimoto}, \citenamefont {Ishikawa}, \citenamefont {O'Farrell},\ and\
  \citenamefont {Nakatsuji}}]{Sagayama2013-im}%
  \BibitemOpen
  \bibfield  {author} {\bibinfo {author} {\bibfnamefont {H.}~\bibnamefont
  {Sagayama}}, \bibinfo {author} {\bibfnamefont {D.}~\bibnamefont {Uematsu}},
  \bibinfo {author} {\bibfnamefont {T.}~\bibnamefont {Arima}}, \bibinfo
  {author} {\bibfnamefont {K.}~\bibnamefont {Sugimoto}}, \bibinfo {author}
  {\bibfnamefont {J.~J.}\ \bibnamefont {Ishikawa}}, \bibinfo {author}
  {\bibfnamefont {E.}~\bibnamefont {O'Farrell}},\ and\ \bibinfo {author}
  {\bibfnamefont {S.}~\bibnamefont {Nakatsuji}},\ }\href
  {https://doi.org/10.1103/physrevb.87.100403} {\bibfield  {journal} {\bibinfo
  {journal} {Phys. Rev. B}\ }\textbf {\bibinfo {volume} {87}},\ \bibinfo
  {pages} {100403} (\bibinfo {year} {2013})}\BibitemShut {NoStop}%
\bibitem [{\citenamefont {Lefrançois}\ \emph {et~al.}(2015)\citenamefont
  {Lefrançois}, \citenamefont {Simonet}, \citenamefont {Ballou}, \citenamefont
  {Lhotel}, \citenamefont {Hadj-Azzem}, \citenamefont {Kodjikian},
  \citenamefont {Lejay}, \citenamefont {Manuel}, \citenamefont {Khalyavin},\
  and\ \citenamefont {Chapon}}]{Lefrancois2015-uz}%
  \BibitemOpen
  \bibfield  {author} {\bibinfo {author} {\bibfnamefont {E.}~\bibnamefont
  {Lefrançois}}, \bibinfo {author} {\bibfnamefont {V.}~\bibnamefont
  {Simonet}}, \bibinfo {author} {\bibfnamefont {R.}~\bibnamefont {Ballou}},
  \bibinfo {author} {\bibfnamefont {E.}~\bibnamefont {Lhotel}}, \bibinfo
  {author} {\bibfnamefont {A.}~\bibnamefont {Hadj-Azzem}}, \bibinfo {author}
  {\bibfnamefont {S.}~\bibnamefont {Kodjikian}}, \bibinfo {author}
  {\bibfnamefont {P.}~\bibnamefont {Lejay}}, \bibinfo {author} {\bibfnamefont
  {P.}~\bibnamefont {Manuel}}, \bibinfo {author} {\bibfnamefont
  {D.}~\bibnamefont {Khalyavin}},\ and\ \bibinfo {author} {\bibfnamefont
  {L.~C.}\ \bibnamefont {Chapon}},\ }\href
  {https://doi.org/10.1103/PhysRevLett.114.247202} {\bibfield  {journal}
  {\bibinfo  {journal} {Phys. Rev. Lett.}\ }\textbf {\bibinfo {volume} {114}},\
  \bibinfo {pages} {247202} (\bibinfo {year} {2015})}\BibitemShut {NoStop}%
\bibitem [{\citenamefont {Donnerer}\ \emph {et~al.}(2016)\citenamefont
  {Donnerer}, \citenamefont {Rahn}, \citenamefont {Sala}, \citenamefont {Vale},
  \citenamefont {Pincini}, \citenamefont {Strempfer}, \citenamefont {Krisch},
  \citenamefont {Prabhakaran}, \citenamefont {Boothroyd},\ and\ \citenamefont
  {McMorrow}}]{Donnerer2016-nx}%
  \BibitemOpen
  \bibfield  {author} {\bibinfo {author} {\bibfnamefont {C.}~\bibnamefont
  {Donnerer}}, \bibinfo {author} {\bibfnamefont {M.~C.}\ \bibnamefont {Rahn}},
  \bibinfo {author} {\bibfnamefont {M.~M.}\ \bibnamefont {Sala}}, \bibinfo
  {author} {\bibfnamefont {J.~G.}\ \bibnamefont {Vale}}, \bibinfo {author}
  {\bibfnamefont {D.}~\bibnamefont {Pincini}}, \bibinfo {author} {\bibfnamefont
  {J.}~\bibnamefont {Strempfer}}, \bibinfo {author} {\bibfnamefont
  {M.}~\bibnamefont {Krisch}}, \bibinfo {author} {\bibfnamefont
  {D.}~\bibnamefont {Prabhakaran}}, \bibinfo {author} {\bibfnamefont {A.~T.}\
  \bibnamefont {Boothroyd}},\ and\ \bibinfo {author} {\bibfnamefont {D.~F.}\
  \bibnamefont {McMorrow}},\ }\href
  {https://doi.org/10.1103/PhysRevLett.117.037201} {\bibfield  {journal}
  {\bibinfo  {journal} {Phys. Rev. Lett.}\ }\textbf {\bibinfo {volume} {117}},\
  \bibinfo {pages} {037201} (\bibinfo {year} {2016})}\BibitemShut {NoStop}%
\bibitem [{\citenamefont {Gallego}\ \emph {et~al.}(2016)\citenamefont
  {Gallego}, \citenamefont {Perez-Mato}, \citenamefont {Elcoro}, \citenamefont
  {Tasci}, \citenamefont {Hanson}, \citenamefont {Aroyo},\ and\ \citenamefont
  {Madariaga}}]{Gallego2016-hq}%
  \BibitemOpen
  \bibfield  {author} {\bibinfo {author} {\bibfnamefont {S.~V.}\ \bibnamefont
  {Gallego}}, \bibinfo {author} {\bibfnamefont {J.~M.}\ \bibnamefont
  {Perez-Mato}}, \bibinfo {author} {\bibfnamefont {L.}~\bibnamefont {Elcoro}},
  \bibinfo {author} {\bibfnamefont {E.~S.}\ \bibnamefont {Tasci}}, \bibinfo
  {author} {\bibfnamefont {R.~M.}\ \bibnamefont {Hanson}}, \bibinfo {author}
  {\bibfnamefont {M.~I.}\ \bibnamefont {Aroyo}},\ and\ \bibinfo {author}
  {\bibfnamefont {G.}~\bibnamefont {Madariaga}},\ }\href
  {https://doi.org/10.1107/S1600576716015491} {\bibfield  {journal} {\bibinfo
  {journal} {J. Appl. Crystallogr.}\ }\textbf {\bibinfo {volume} {49}},\
  \bibinfo {pages} {1941} (\bibinfo {year} {2016})}\BibitemShut {NoStop}%
\bibitem [{\citenamefont {Gražulis}\ \emph {et~al.}(2012)\citenamefont
  {Gražulis}, \citenamefont {Daškevič}, \citenamefont {Merkys},
  \citenamefont {Chateigner}, \citenamefont {Lutterotti}, \citenamefont
  {Quirós}, \citenamefont {Serebryanaya}, \citenamefont {Moeck}, \citenamefont
  {Downs},\ and\ \citenamefont {Le~Bail}}]{Grazulis2012-mb}%
  \BibitemOpen
  \bibfield  {author} {\bibinfo {author} {\bibfnamefont {S.}~\bibnamefont
  {Gražulis}}, \bibinfo {author} {\bibfnamefont {A.}~\bibnamefont
  {Daškevič}}, \bibinfo {author} {\bibfnamefont {A.}~\bibnamefont {Merkys}},
  \bibinfo {author} {\bibfnamefont {D.}~\bibnamefont {Chateigner}}, \bibinfo
  {author} {\bibfnamefont {L.}~\bibnamefont {Lutterotti}}, \bibinfo {author}
  {\bibfnamefont {M.}~\bibnamefont {Quirós}}, \bibinfo {author} {\bibfnamefont
  {N.~R.}\ \bibnamefont {Serebryanaya}}, \bibinfo {author} {\bibfnamefont
  {P.}~\bibnamefont {Moeck}}, \bibinfo {author} {\bibfnamefont {R.~T.}\
  \bibnamefont {Downs}},\ and\ \bibinfo {author} {\bibfnamefont
  {A.}~\bibnamefont {Le~Bail}},\ }\href {https://doi.org/10.1093/nar/gkr900}
  {\bibfield  {journal} {\bibinfo  {journal} {Nucleic Acids Res.}\ }\textbf
  {\bibinfo {volume} {40}},\ \bibinfo {pages} {D420} (\bibinfo {year}
  {2012})}\BibitemShut {NoStop}%
\end{thebibliography}
